\documentclass[final,4p,times,twocolumn]{elsarticle}

\usepackage{amssymb}
\usepackage{amsmath}

\newcommand{\avgb}{\langle B \rangle}
\newcommand{\avgnpart}{\langle N_{\rm part} \rangle}

\journal{Nuclear Physics B}

\begin{document}

\begin{frontmatter}

\title{The role of strangeness in baryon and electric charge stoppings}

\author{Zi-Wei Lin, Mason Alexander Ross} 

\affiliation{organization={Department of Physics, East Carolina University},
            addressline={C-209 Howell Science Complex}, 
            city={Greenville},
            postcode={27858}, 
            state={NC},
            country={USA}}

\begin{abstract}
Recently it has been proposed that comparing the net-baryon ($B$)
stopping with net-electric charge ($Q$) stopping can help studies of
the baryon stopping mechanism in nuclear collisions. 
Here we find the $B/Q\times Z/A$ ratio to be very sensitive to the
difference between $s$ and $\bar s$ quark rapidity distributions. 
For mid-rapidity of isobar collisions at 200A GeV,  a
multi-phase transport (AMPT) model gives slightly more $\bar s$ than
$s$, which leads to $B/Q\times Z/A<1$, while the model  without the
$s-\bar s$ asymmetry gives $B/Q\times Z/A \geq 1$.
Comparing Ru+Ru and Zr+Zr isobar collisions, the AMPT (and UrQMD) 
model gives $B/\Delta Q\times \Delta Z/A<1$ at mid-rapidity at all
centralities, which strongly contradicts the recent STAR data. We also
find that the $B/\Delta  Q\times \Delta Z/A$ ratio is 
very sensitive to the net-light quark ($u,d$) stoppings, but it is
less sensitive to the $s-\bar s$ asymmetry than the $B/Q\times Z/A$
ratio by a factor of 3.  These results are expected to help us resolve
the stopping puzzle  and better understand the baryon stopping
mechanism in the future. 
\end{abstract}


\end{frontmatter}


\section{Introduction}

How much the incoming nucleons lose energy right after a nuclear
collision, which can be called the net-baryon stopping, determines the
initial energy density and net-baryon density that will affect the
equation of state and the evolution of the created dense matter. 
However, baryon stopping is not well understood either qualitatively
or quantitatively. In the standard model, each quark carries a baryon
number of 1/3 while each antiquark carries -1/3, and a baryon consists
of three valence quarks that are the carriers of both the net-baryon
number and net-electric charge. On the other hand, the baryon junction
model sees the baryon as being composed of three quarks connected by a
Y-shaped string junction.

Recently the ratio $B/Q\times Z/A$ has been proposed~\cite{Xu:2022},
where $B$ and $Q$ respectively represent the net-baryon and   
net-electric charge  within a given acceptance, and $Z$ and $A$ are
respectively  the atomic number and mass number of the incoming
nucleus in symmetric A+A collisions. The naive expectation is
$B/Q\times Z/A=1$~\cite{Lin:2024}, because this is true for
the full phase space due to the corresponding conservation laws 
when neglecting the neutron skin effect~\cite{Ross:2025qxr}.  
As the net-electric charge is very difficult to measure
experimentally, the ratio $B/\Delta Q\times \Delta
Z/A$ for Zr+Zr and Ru+Ru isobar collisions at RHIC has been measured
instead~\cite{STAR:2024lvy}; this is possible because of the
large statistics and cancellation of certain systematic errors. 
In our study~\cite{Ross:2025qxr} summarized below, 
we examine the two ratios in terms of quark distributions and
demonstrate the role of different quark flavors including
strangeness. We also present results from the the string-melting AMPT
model including a modified version where the initial $s-\bar s$
asymmetry is removed.

\section{The $B/Q\times Z/A$ ratio in nuclear collisions}

Let us consider the parton phase created in high energy nuclear
collisions under the conventional picture where the baryon and
electric charges are carried by quarks and antiquarks. 
Denoting the rapidity distribution $dN/dy$ of quark flavor $i$ as
$f_i$, we can write
\begin{eqnarray}
f_B \equiv dN_B/dy =(f_u-f_{\bar u}+f_d-f_{\bar d}+f_s-f_{\bar s})/3, 
\nonumber \\
f_Q \equiv dN_Q/dy =(2f_u-2f_{\bar
    u}-f_d+f_{\bar d} -f_s+f_{\bar s})/3. 
\label{bq}
\end{eqnarray}
Thus at a given rapidity $y$, $B/Q\times Z/A \equiv f_B/f_Q\times Z/A$ 
depends on several variables, because in general the quark rapidity
distributions can all be different.

A useful limit is the isospin-symmetric case when $A=2Z$, where we
expect (when neglecting the neutron skin effect) $f_u=f_d \equiv f_q$
and  $f_{\bar u}=f_{\bar d} \equiv f_{\bar q}$, we then have
\begin{eqnarray}
B & \equiv f_B=( 2f_q-2f_{\bar q}+f_s-f_{\bar s} )/3,\nonumber \\
Q & \equiv  f_Q=(f_q-f_{\bar q}-f_s+f_{\bar s})/3.  
\label{bqSym}
\end{eqnarray}
If $f_s=f_{\bar s}$ (i.e., strange and anti-strange quarks have
the same rapidity distribution), we then get
\begin{eqnarray}
B/Q\times Z/A=1  {\rm~at~any~}y,
\label{bqSym2}
\end{eqnarray}
regardless of the stopping of net-light quarks (meaning $u,d$ in this
study). Equivalently, if $B/Q\times Z/A \neq 1$ at a certain
rapidity  for isospin-symmetric A+A collisions, this 
then indicates the  $s-\bar s$ asymmetry. 
Note that the above analysis may also apply to the gluon junction
picture because a gluon junction must connect three quarks to
represent a baryon. Also note that the neutron skin 
effect will modify some of the above relations. Nevertheless, the
above analysis clearly demonstrates the important role that
strangeness play in the $B/Q\times Z/A$ ratio.

\begin{figure}[htb]
\centering
\includegraphics[width=1.0\linewidth]{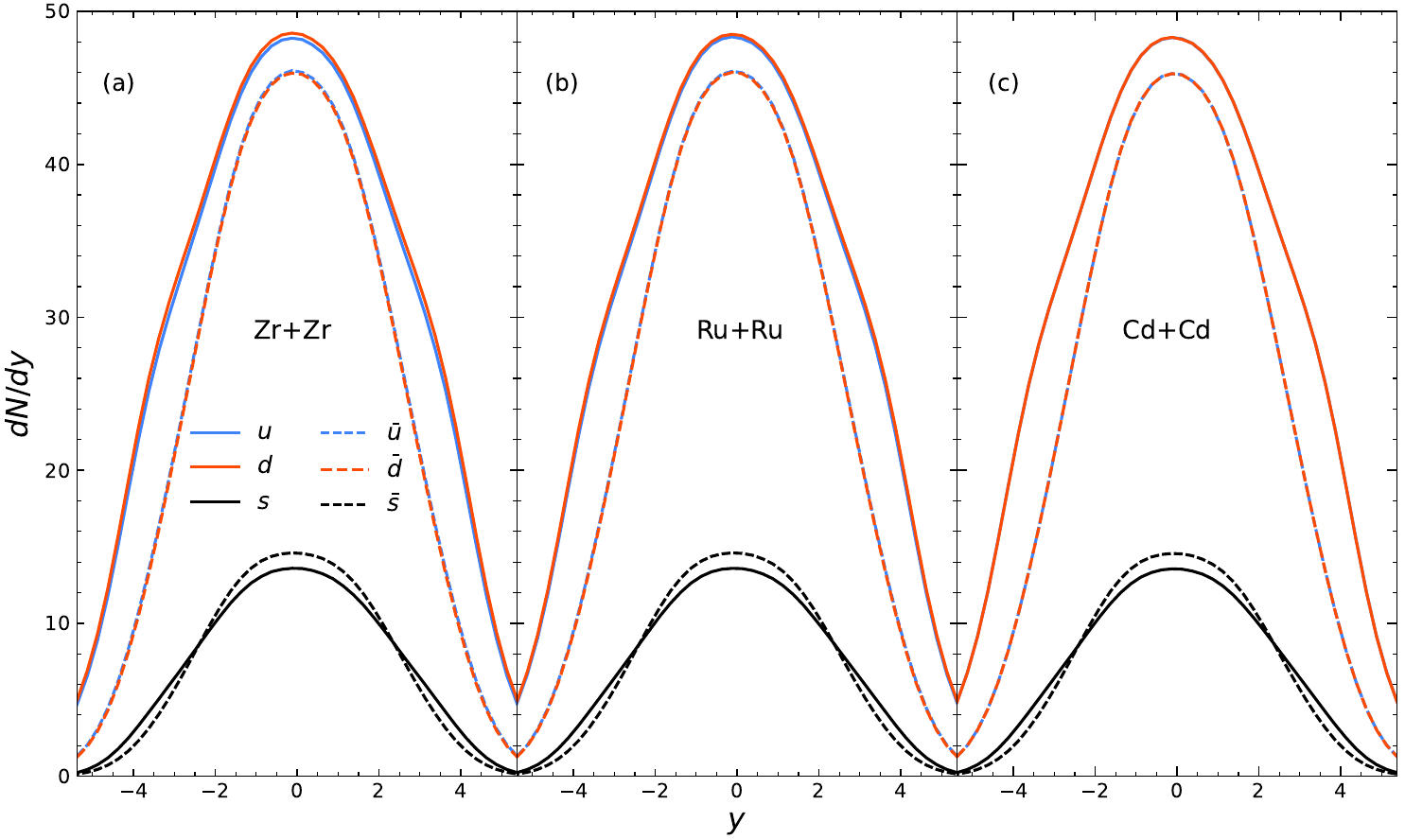}
\caption{Rapidity distributions of initial quarks and antiquarks
  (i.e., before parton scatterings) from the AMPT model for minimum
  bias (a) Zr+Zr, (b) Ru+Ru, and (c) Cd+Cd isobar collisions at $200A$
  GeV.}
\label{figdndy}
\end{figure}

We show in Fig.~\ref{figdndy} the rapidity distributions of initial quarks
in Zr+Zr, Ru+Ru and the hypothetical Cd+Cd isobar 
collisions (at $A=96$) from the AMPT model, where initial quarks refer
to those before the parton cascade. We see a small but obvious
difference between strange and anti-strange quark rapidity
distributions, which is similar for the three collision systems.  Around
mid-rapidity ($|y| \lesssim 2$) there are more anti-strange than
strange quarks while the opposite happens at large rapidities.  
Around mid-rapidity, this $s-\bar s$ asymmetry gives a negative
contribution to $B$ but a positive contribution to $Q$, as shown in
Eq.\eqref{bq}; thus the asymmetry suppresses the $B/Q\times Z/A$
ratio. For the isospin-symmetric $^{96}$Cd system, we see that
$f_u=f_d$ and $f_{\bar u}=f_{\bar d}$ as expected, thus the $s-\bar s$
asymmetry (i.e., $f_s \neq  f_{\bar s}$) is the only reason for
$B/Q\times Z/A \neq 1$ for Cd+Cd collisions according to
Eqs.~\eqref{bqSym}-\eqref{bqSym2}.

\begin{figure}[htb]
\centering
\includegraphics[width=1.0\linewidth]{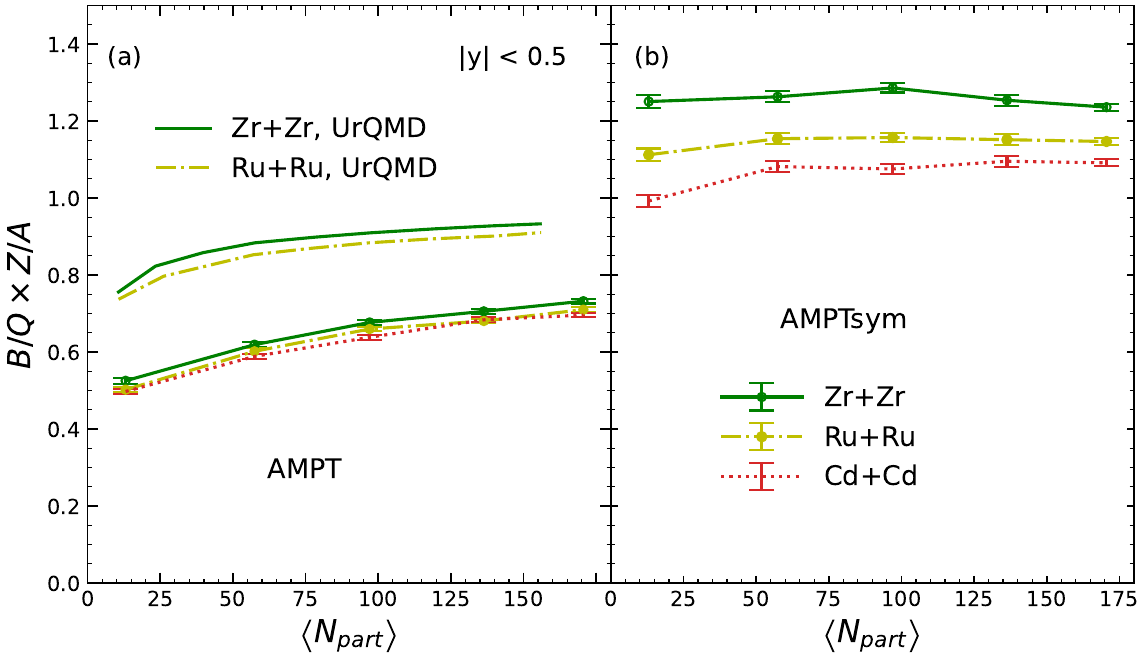}
\caption{The $B/Q\times Z/A$ ratio as a function of $\avgnpart$ for
  isobar collisions from (a) the (normal) AMPT model  and the UrQMD model, (b) the
  modified AMPT model (with symmetrized initial $s$ and $\bar s$
  distributions).}
\label{figbqza}
\end{figure}

Figure~\ref{figbqza} shows the hadron $B/Q\times Z/A$ ratio at
mid-rapidity as a function of $\avgnpart$ (the average number of
participant nucleons) in isobar collisions from the AMPT and UrQMD 
models. We see  in panel (a) that the ratios from the normal AMPT model
all gradually increase with $\avgnpart$ but stay below unity.   
Results from the UrQMD model~\cite{Lewis:2022arg} also show these
features, although they are higher. When we remove the $s-\bar 
s$ asymmetry by symmetrizing the initial energy-momentum and
space-time distributions of $s$ and $\bar s$ quarks before the parton
cascade (the AMPTsym model)~\cite{Ross:2025qxr}, the $B/Q\times Z/A$
ratios at mid-rapidity in panel (b) are mostly above one. Therefore,
the $s-\bar s$ asymmetry in the initial rapidity distribution, which
could come from the Lund string fragmentation, strongly affects the 
$B/Q\times Z/A$ ratio in nuclear collisions. We also see in
Fig.~\ref{figbqza} that the $B/Q\times Z/A$ ratio at mid-rapidity
depends on the isospin asymmetry of the incoming 
nuclei.  The ordering is also the same for all three models (AMPT,
AMPTsym, and UrQMD), with the highest value for Zr+Zr collisions and
the next highest value for Ru+Ru collisions (at the same $\avgnpart$).
We find that this is also the case for initial partons from the two
versions of the AMPT model~\cite{Ross:2025qxr}. The net-baryon
distributions are essentially the same for the three  isobar
collisions, thus this ordering results from the net-electric 
charge $Q$. The separate effects of parton cascade, quark coalescence
hadronization, and hadron cascade have also been 
investigated~\cite{Ross:2025qxr}. 

To see how a small $s-\bar s$ asymmetry gives a
large deviation of $B/Q\times Z/A$ from unity, we rewrite
Eq.\eqref{bq} as  
\begin{eqnarray}  
B&=(f_{net-u}+f_{net-d}+f_{net-s})/3,\nonumber \\
Q&=(2f_{net-u}-f_{net-d}-f_{net-s})/3,  
\label{bq2}
\end{eqnarray}  
where $f_{net-u} \equiv f_u-f_{\bar u}$ represents the net-$u$ quark
stopping. For collisions of most nuclei (where $Z \sim A/2$), we get 
\begin{eqnarray} 
B/Q\times Z/A \!\!\!\!\!\!  &\simeq& \!\!\!\!\!\!   \frac{Z}{A} \left
                (\frac{f_{net-u}+f_{net-d}}{2f_{net-u}-f_{net-d}}
                \right   ) \left (1+\frac{3f_{net-s}}{2f_{net-q}} 
                \right ) \\
&{\rm with~}& f_{net-q} \equiv (f_{net-u}+f_{net-d})/2.\nonumber 
\label{bqza} 
\end{eqnarray}
For example, for isospin symmetric A+A collisions,
$f_{net-u}=f_{net-d}$ and thus the above reduces to
\begin{eqnarray}
B/Q\times Z/A \simeq 1+\frac{ 3}{2}\frac{f_{net-s}}{f_{net-q}}.
\end{eqnarray}
Although the $s-\bar s$ asymmetry (i.e., $f_{net-s}/f_s$) is quite
small ($\lesssim 10\%$) as shown in Fig.~\ref{figdndy},
the $f_{net-s}/f_{net-q}$ term above is much higher and thus causes a
big deviation ($\sim 40\%$) of $B/Q\times Z/A$ from unity at
mid-rapidity.

\section{The $B/\Delta Q\times \Delta Z/A$ ratio for isobar
  collisions} 

\begin{figure}[htb]
\centering
\includegraphics[width=1.0\linewidth]{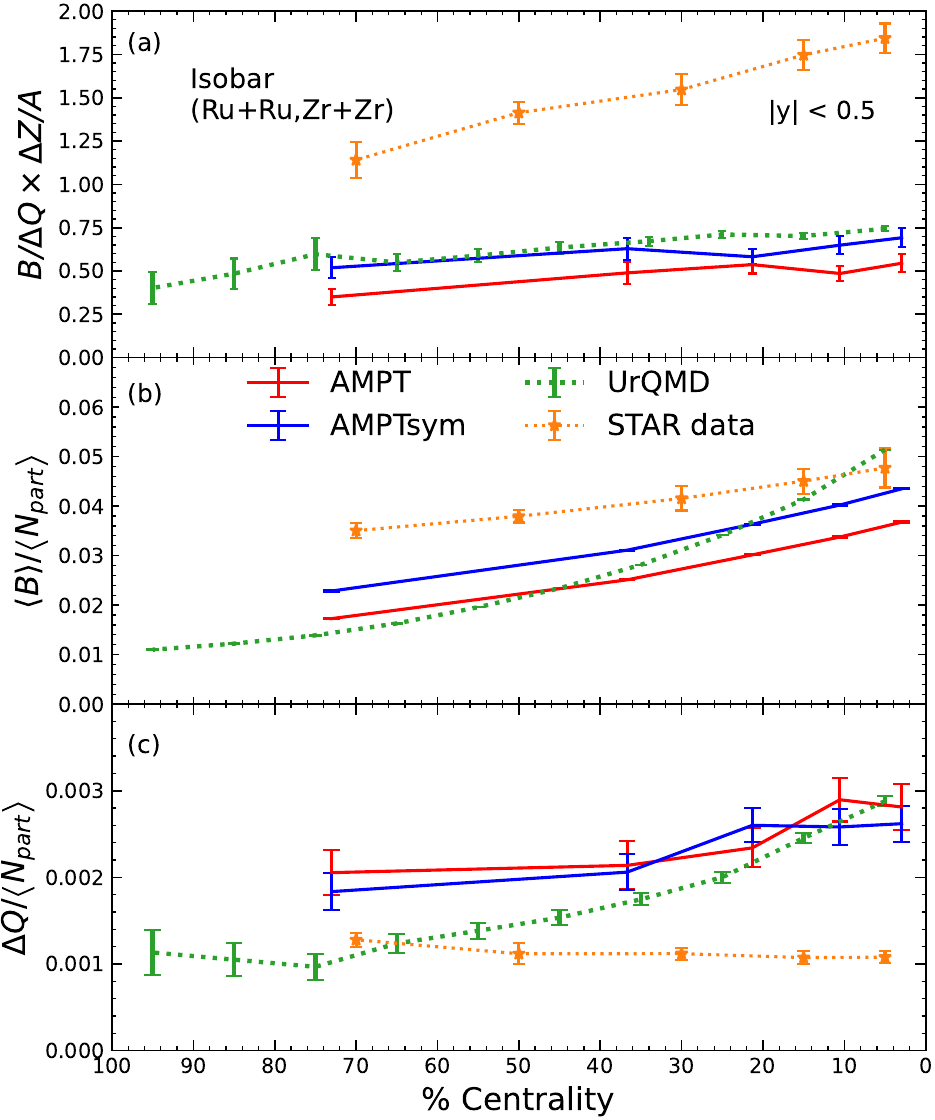} 
\caption{Mid-rapidity (a) $B/\Delta Q\times \Delta Z/A$, 
(b) average net-baryon number (scaled by $\avgnpart$) and (c)
net-electric charge difference (scaled by $\avgnpart$) versus
centrality for isobar  collisions from the AMPT and UrQMD models in
comparisons with the STAR data with statistical error
bars~\cite{STAR:2024lvy}.} 
\label{figisobar}
\end{figure}

We show in Fig.~\ref{figisobar} the hadron results at mid-rapidity for
Ru+Ru and Zr+Zr isobar collisions at 200A GeV, where the difference
in a variable $x$ is $\Delta x \equiv x_{\rm Ru+Ru}-x_{\rm
  Zr+Zr}$. We see that the $B/\Delta Q\times \Delta Z/A$ ratios from
all three models are well below unity, where the ratios from the
AMPTsym model are close to those from the UrQMD model. 
As expected, the $\avgb/\avgnpart$ values from the normal AMPT 
model are lower than those in the AMPTsym model due to the $s-\bar s$
asymmetry as shown in Fig.~\ref{figisobar}(b). 
We also see in Fig.~\ref{figisobar}(c) that the two AMPT models have
almost the same $\Delta Q/\avgnpart$. 
Overall, we see that the AMPT results on
$\avgb/\avgnpart$ and  $\Delta Q/\avgnpart$ are different from the
UrQMD results while they overlap at certain centralities. 
On the other hand, the $B/\Delta Q\times \Delta Z/A$ STAR
data~\cite{STAR:2024lvy} at mid-rapidity  are much higher (mostly by a
factor of 2 or more) than all the model results shown in
Fig.~\ref{figisobar}(a) at all centralities. 
Therefore, we have a puzzle about the baryon versus charge stoppings in
Ru+Ru and Zr+Zr isobar collisions, despite the uncertainty in the
initial $s-\bar s$ asymmetry.  Note that the UrQMD results
shown in Fig.~\ref{figisobar} use the UrQMD $\avgnpart$
values~\cite{Lv:2025} and are thus not the same as the UrQMD curves
shown in the recent STAR study~\cite{STAR:2024lvy} 
(where the experimental $\avgnpart$ values were used).

\begin{figure}[htb]
\centering
\includegraphics[width=1.0\linewidth]{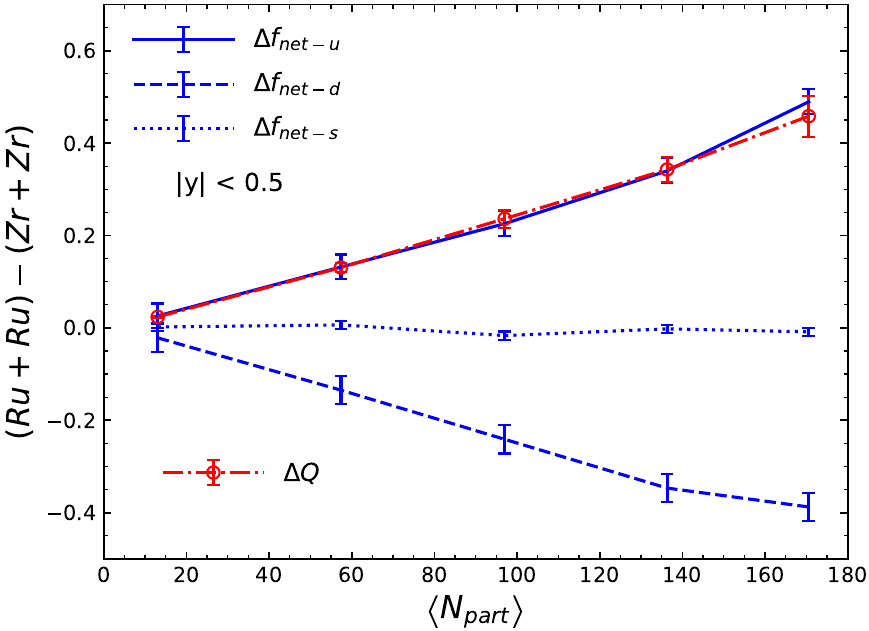}
\caption{The differences of the net-quark numbers of different flavors
  and net-electric charge for isobar collisions at mid-rapidity as
  functions of $\avgnpart$ from the normal AMPT model.} 
\label{figdiff}
\end{figure}

To further understand these results, let us rewrite Eq.\eqref{bq2} for
the differences between two isobar systems as
\begin{eqnarray}
\Delta B&=(\Delta f_{net-u}+\Delta f_{net-d}+\Delta f_{net-s})/3,
          \nonumber \\
\Delta Q&=(2\Delta f_{net-u}-\Delta f_{net-d}-\Delta f_{net-s})/3,\\
&{\rm with~} \Delta f_{net-u} = \Delta f_u -\Delta f_{\bar u}.\nonumber 
\end{eqnarray}
Results from the normal AMPT model in Fig.~\ref{figdiff} show $\Delta
f_{net-s} \simeq 0$. Coupled with the observation $\Delta B \simeq 0$
for isobar systems, we then have 
\begin{equation}
\Delta f_{net-u} \simeq -\Delta f_{net-d}, 
~~\Delta Q \simeq \Delta f_{net-u}, 
\end{equation}
both of which are confirmed by the AMPT results in Fig.~\ref{figdiff}. 
Note that these relations are also valid for the differences between
Cd+Cd and Ru+Ru isobar collisions~\cite{Ross:2025qxr}. 
At the parton level, we then get  
\begin{eqnarray} 
B/\Delta Q \! \times \! \Delta Z/A 
\!\! \simeq \!\! \frac{\Delta Z~B}{A\Delta f_{net-u}}
\!\! \simeq \!\! \frac{\Delta Z}{A} \! \left (
  \!\frac{2f_{net-q}}{\Delta f_{net-u}} \! \right ) \left ( 
  \! 1 \!\!+\!\frac{f_{net-s}}{2f_{net-q}} \! \right ) \! . 
\label{bdq}
\end{eqnarray}
Comparing the above with Eq.\eqref{bqza}, we see that
the $B/Q\times Z/A$ ratio is more sensitive to $f_{net-s}$ (or the 
$s-\bar s$ asymmetry) by a factor of 3 than the $B/\Delta Q\times
\Delta Z/A$ ratio. 
On the other hand, the $B/\Delta Q\times
\Delta Z/A$ ratio is mostly sensitive to the net-light quark ($u,d$) 
stoppings ($f_{net-q}$) and their change between the two isobar
systems ($\Delta f_{net-u}$).  

\section{Summary}

Baryon stopping in nuclear collisions has not been well
understood. Recently it has been proposed that comparing the
net-baryon with the net-electric charge in nuclear collisions can
help. Here we show that the $B/Q\times Z/A$ ratio can strongly depend
on rapidity, and it also depends sensitively on the difference between 
$s$ and $\bar s$ rapidity distributions at the quark level. 
The AMPT model gives several percent more $\bar s$ than $s$ at
mid-rapidity, which leads to a $B/Q\times Z/A$ ratio well below unity
for Zr+Zr and Ru+Ru isobar collisions at 200A GeV.   
For collisions of isospin symmetric nuclei such as $^{40}\rm Ca$, the
$B/Q\times Z/A$ ratio would deviate from unity only because of this
$s-\bar s$ asymmetry (when the neutron skin effect is neglected).  
Note that including the neutron skin or realistic nuclear density profiles,
which is left for future work, could lead to modifications of the
centrality dependence of electric charge stopping and consequently the
baryon-to-charge ratios. 

We also find that the $B/\Delta Q\times \Delta Z/A$ ratio
depends crucially on the net-light quark ($u,d$) stoppings, 
but it is less sensitive to the $s-\bar s$ asymmetry than the $B/Q\times
Z/A$ ratio by a factor of 3. For isobar collisions, the AMPT model
gives $B/\Delta Q\times \Delta Z/A <1$ (regardless of the $s-\bar s$
asymmetry); this is similar to results from the UrQMD model but
strongly contradicts the recent data from the STAR Collaboration. 
The dependences of the baryon-to-electric charge ratios on the
stoppings of different quark flavors obtained from this study 
will be useful for resolving the large disagreements between
theoretical models and the experimental data and deepen our
understanding of baryon and electric charge stoppings. 

\section*{Acknowledgments}
This work has been supported by the National Science Foundation under
Grant No. 2310021.


\begin{thebibliography}{00}

\bibitem{Xu:2022} 
Z. Xu, at the RBRC Workshop on Physics Opportunities from Isobar Run, https://in-
dico.bnl.gov/event/13769, 2022.

\bibitem{Lin:2024}
Z.~W.~Lin, at 1st workshop on Baryon Dynamics from RHIC to EIC, Stony Brook
University, https://indico.cfnssbu.physics.sunysb.edu/event/113/, 2024.

\bibitem{Ross:2025qxr}
M.~A.~Ross and Z.~W.~Lin,
Eur. Phys. J. C \textbf{86}, 143 (2026).

\bibitem{STAR:2024lvy}
STAR, 
Science \textbf{393}, 727 (2026).

\bibitem{Lewis:2022arg}
N.~Lewis, W.~Lv, M.~A.~Ross, C.~Y.~Tsang, J.~D.~Brandenburg, Z.~W.~Lin, R.~Ma, Z.~Tang, P.~Tribedy and Z.~Xu,
Eur. Phys. J. C \textbf{84}, 590 (2024).

\bibitem{Lv:2025}
W. Lv, private communications regarding UrQMD simulations, 2025.


\end{thebibliography}
\end{document}